\documentclass[a4paper,11pt]{article}
\pdfoutput=1
\usepackage{amssymb,amsmath,amsfonts,makeidx,placeins,pbox,multirow}
\usepackage{graphicx,rotate,subcaption,color,slashed,cite,caption,epstopdf,verbatim}
\usepackage{longtable,tabu}
\usepackage{array}
\usepackage{color}
\usepackage{mathtools}
\usepackage[normalem]{ulem}
\newcolumntype{P}[1]{>{\centering\arraybackslash}p{#1}}
\usepackage[colorlinks=true,
            linkcolor=magenta,
            urlcolor=blue,
            citecolor=blue]{hyperref}
\evensidemargin=\oddsidemargin
\def\bar {\overline}

\def\bea {\begin{eqnarray}}
\def\eea {\end{eqnarray}}

\def\beq{\begin{equation}}
\def\eeq{\end{equation}}
\def\barr{\begin{array}}
\def\earr{\end{array}}

\def\gev{\ensuremath{\mathrm{Ge\kern -0.1em V}}}

\def\beq{\begin{equation}}
\def\eeq{\end{equation}}
\def\bea{\begin{eqnarray}}
\def\eea{\end{eqnarray}}
\def\bmat{\begin{pmatrix}}
\def\emat{\end{pmatrix}}
\def\to{\rightarrow}

\begin{document}
\begin{center}
{\Large \bf Emergent 2HDM in LSS Little-Higgs: Musings from Flavor and} \\
\vspace{5pt}
{\Large \bf Electroweak Physics} \\
\vspace*{0.5cm} {\sf  $^{a)}$ Nilanjana Kumar\footnote{nilanjana.kumar@gmail.com}, \ $^{a)}$ $^{b)}$Soumya Sadhukhan~\footnote{physicsoumya@gmail.com }} \\
\vspace{10pt} {\small } {\em $^{a)}$ Department of Physics and Astrophysics, University of Delhi, New Delhi, 110007, India\\
$^{b)}$ Ramakrishna Mission Residential College (Autonomous), Narendrapur, Kolkata 700103, India} \\
\normalsize
\end{center}
\bigskip
\begin{abstract}
The low energy effective theory ($\sim$ TeV) of the little-Higgs model with $SU(6)/Sp(6)$, as proposed by 
Low, Skiba and Smith (LSS), exhibits a
two-Higgs doublet model (2HDM) structure. 
The symmetry dictates interesting Yukawa patterns, translating to non-trivial fermion couplings with 
both of the Higgs doublets. 
The couplings of the scalars with the fermions can 
induce flavor changing neutral currents (FCNC), 
which get constraints from flavor physics observables such as BR$(B\rightarrow X_s\gamma)$, $B_s - \bar{B}_s$ mixing etc.
The precision measurement of $Z b \bar{b}$ vertex, the top and Higgs mass along with other Higgs coupling measurements 
at the Large Hadron Collider (LHC) also enforce severe restrictions on the LSS model. 
Direct LHC search results of beyond the Standard Model (BSM) particles also impose bounds on the masses.
We probe the LSS model in view of the above constraints through a random scan in the multi-dimensional parameter space.
We observe, 
on contrary to the general 2HDM scenario, the emergent 2HDM from the LSS model is less constrained 
from the flavor data and the $Z b \bar{b}$ measurement
but is severely constrained form the electroweak (EW) searches at the LHC. From the flavor data and $Z b \bar{b}$,
we find that the charged Higgs mass is relaxed with $\tan\beta$ being restricted to $0.5-5$, whereas
the charged Higgs mass is pushed to larger than 1 TeV along with $\tan\beta$ being further restricted to $< 3$ when the LHC bounds are incorporated. 
\end{abstract}

\section{Introduction}
Even if the Standard Model (SM) is the favorite candidate to explain all the 
results obtained at the LHC, models with extended symmetries 
beyond the SM are of great interest due to their elegant UV completion 
along with the power of stabilizing the scalars against radiative corrections, i.e. 
solving the gauge hierarchy problem~\cite{PhysRevD.14.1667}. 
There are different broad classes 
of models with enhanced symmetries. Firstly, the models where the space-time symmetry 
is enhanced to incorporate the spin, also to have symmetries that even 
relate the bosons and the fermions; are called supersymmetric models~\cite{Fayet:1976et}. 
Another set of models also possess enhanced gauge symmetries which are
systematically broken, sometimes in multiple steps, to emerge the Higgs as a pseudo Nambu-Goldstone boson (pNGB). 
In these models, such as little Higgs~\cite{ArkaniHamed:2001nc} and composite Higgs~\cite{Kaplan:1983fs} scenarios, the extended scalar sector manifests from the Goldstone bosons and unbroken symmetry does 
contain the SM electro-weak group $SU(2)\times U(1)$.
In these sets of models, where the Higgs boson emerges as a ``little'' part of a bigger representation and when some global 
symmetry is broken by the interplay between two or more coupling constants, are termed 
as little Higgs models (LH)~\cite{ArkaniHamed:2001nc} and many different variations 
of this model have been proposed in literature~\cite{Cheng:2004yc,Cheng:2003ju,Low:2004xc,Kaplan:2003uc}. 
This {\it collective symmetry breaking} is the essential ingredient in the little Higgs theories in 
order to free the Higgs mass-squared from quadratic divergences at one loop.
On the other hand, in some models, when the enlarged global symmetry is broken by some strong 
dynamics and the Higgs emerges as the pseudo Nambu-Goldstone boson of that enlarged global symmetry, 
is termed composite Higgs models~\cite{Contino:2010rs,Agashe:2004rs,Panico:2012uw,Berger:2012ec}.
Here, one has to assume partial compositeness ~\cite{Sannino:2016sfx} of the SM fermions in order to 
generate their masses.

Due to the presence of the Nambu Goldstone bosons (NGB) in the little Higgs 
theories~\cite{Schmaltz:2005ky,Perelstein:2005ka} and the composite Higgs models~\cite{Ma:2015gra,Cai:2018tet,Bertuzzo:2012ya,Mrazek:2011iu}, it is natural that they can usher in an extended scalar sector at TeV scale. 
Some scenarios among these models represent an effective 2HDM~\cite{Branco:2011iw}
framework at the lower energy scale ($\sim$ TeV). We take particular interest to study 
the case where the global symmetry $SU(6)$ is broken to an $Sp(6)$ to provide an extended LH group structure $SU(6)/Sp(6)$, which is one variation of the little Higgs scenario developed by Low, Skiba and Smith (LSS Model)~\cite{Low:2002ws}.
 Similarly, there are composite Higgs models based on the same group structure $SU(6)/Sp(6)$~\cite{Ma:2015gra,Cai:2018tet}, where, similar to the little Higgs model,
the low-energy effective theory exhibits a 2HDM structure and in addition, beyond standard model (BSM) 
particles exist. Even though the structure of these two models are alike, there 
are some differences between these two sets of models.
The gauged subgroup in the $SU(6)/Sp(6)$
little Higgs model is more extended than the electroweak (EW) subgroup and the additional gauge generators
are obtained via the symmetry breaking at the condensation scale ($f$). Also, the technical difference
in the little Higgs and composite Higgs scenario is the generation of the scalar potential terms with different quartic terms due to the non-identical Yukawa structures of these two models. Hence, the dependencies of the emergent 2HDM
scalar potential coefficients on different set of model parameters are not similar in both
cases. In this work, we choose to explore the LSS model with an emergent 2HDM structure 
at low energy in the view of recent flavor and LHC data.

In addition to the extended scalar sector, the BSM sector of the LSS model includes new fermions, such as one vector-like quark doublet and two vector-like quark singlets 
along with new gauge bosons. 
Natural realization of the LSS model in the low energy effective theory 
can be executed through the alignment limit of the emergent 2HDM~\cite{Low:2002ws,Gopalakrishna:2015dkt}.
In the familiar 2HDM~\cite{Branco:2011iw}, the Yukawa sector is organized arbitrarily using the $Z_2$ symmetry, 
according to the phenomenological requirements, whereas, in the little Higgs models, a larger symmetry dictates it. 
On the other hand, in the LSS scenario, the Yukawa sector which is initially arranged in terms of the fermionic multiplets aka sextets, directs the pattern of the Yukawa sector in emergent 2HDM after the symmetry 
breaking. 
Moreover, ultraviolet completion of the 2HDM is possible through the LSS model with an 
extended gauge sector $SU(6)/Sp(6)$. Therefore, a study of this model 
can reveal the effects of larger symmetry on the experimental observables, that 
are unlikely to manifest in {\it ad hoc} generic 2HDM.
A comprehensive phenomenological analysis of the LSS model 
is shown in Ref:~\cite{Gopalakrishna:2015dkt}, considering  
8 TeV LHC results. But the extended scalar sector of the LSS model demands a
detailed study of collider phenomenology as well as the effects of BSM scalars in flavor physics
observables.
In this paper, we discuss the effective 2HDM emerging from the LSS model at low energy 
in the context of flavor and electroweak (EW) observables, along with constraints from 13 TeV LHC.  

In light of the significant tension between the SM and experimental 
measurements of lepton flavor universality (LFU) observables~\cite{Amhis:2016xyh}, it is worthwhile to look for a 
NP model, capable of explaining the discrepancies. 
In the LH models, the presence of new particles with possible non-trivial flavor structure 
can potentially address the challenges in the flavor sector.
The flavor structure of different little Higgs models had been studied in the literature
through multiple variants~\cite{delAguila:2019htj,Fajfer:2007vk,Bigi:2009df},
along with some phenomenological studies~\cite{Han:2003wu}.
The major focus of these models had been to probe the effects of vectorlike fermions and the gauge bosons,
significantly modifying the flavor observables. 
The SM extensions with two Higgs doublets, such as Refs:~\cite{Mahmoudi:2009zx,PhysRevD.41.3421} and various other scalar extensions
of the SM~\cite{Haber:1999zh,Banerjee:2019gmr} can also address the 
flavor anomalies with the presence of the extra scalars in the theory.  
This motivates us to particularly investigate the effect of the additional 
scalars of the LSS model in the flavor sector.
It is worthwhile to study how the neutral and the charged BSM scalars 
can modify the flavor and other EW observables and therefore impose 
constraints on the emergent 2HDM parameters derived from the LSS little Higgs model. 
%
At first, we assess the impact of different flavor observables and the $Zb\bar b$ vertex correction. 
Amongst all BSM scalars present, we take particular interest to study the couplings of the charged Higgs to the 
third generation quarks, to estimate the modifications in specific flavor observables. 
We observe that the $B$-meson mixing, $B$-meson decays and correction to the $Zb\bar{b}$ vertex impose
important constraints on the model parameter space. 

In the LSS little-Higgs model, the lightest
CP-even neutral state is identified with 125 GeV Higgs, observed at the LHC. 
Precise measurements of top and bottom mass along with other measurements in the Higgs sector place non-trivial constraints
on the model parameter space. 
Non-observation of new states at the LHC~\cite{Buckley:2020wzk,RAPPOCCIO2019100027}
and the EW precision tests (EWPT)~\cite{ALEPH:2005ab} place strong 
limits on the BSM sector and hence also constrain the LSS model.
We determine the allowed regions of $\tan \beta$ and 
charged Higgs mass in the emergent 2HDM structure of the LSS model as compared to a generic 2HDM and study 
the crucial phenomenological differences.
In contrast to the generic 2HDM with specific Yukawa structure 
(type-I,II,III 2HDM etc), where the flavor constraints 
play a major role in ruling out a large parameter space, in the emergent 2HDM of the LSS model, 
constraints from EW searches at the LHC are found to be more severe than those coming from the flavor and $Zb\bar b$. 
This is due to the predictive 
nature of the Yukawa sector in the LSS model that shrinks the parameter 
space mainly to fix the Higgs mass, top quark mass and top quark Yukawa coupling. 

In section~\ref{theory}, we describe the Yukawa interaction in the LSS model to obtain the 
couplings of the scalar to the SM and BSM fermions. Then we show how an effective 2HDM framework 
can be worked out in the LSS model, connecting the strong sector parameters of the LSS model
to the entities of the emergent 2HDM. We also show how different Yukawa structures can be 
arranged in some limits of the bottom quark Yukawa couplings. In section~\ref{constrains}, 
we focus on the different constraints from flavor and EW observables in the 2HDM framework of LSS model. 
First, we consider all the 
relevant flavor and EW observables, discussing their possible origin and effects. 
Then, we scan over the entire model parameter space of the LSS model, 
curving out the region favored by the flavor and EW observables. 
Next, we bring in the LHC measurements and discuss 
their super-constraining effects on the parameter space, which is already constrained from flavor and EW data. 
Finally, in section~\ref{conclusion}, we chart out relative importance of our results in comparison to a generic 2HDM.

\section{Low-Skiba-Smith (LSS) Little Higgs Model}
\label{theory}
The little Higgs model proposed by Low, Skiba and Smith i.e. the LSS model~\cite{Low:2002ws}
exhibits a larger symmetry $SU(6)$ in the unbroken form, which is broken to a 
residual symmetry $Sp(6)$ by the field condensates. 
The number of massless Goldstone bosons that are expected to appear due 
to the symmetry breaking, is equal to the difference in the number of 
generators, i.e. $35-21 = 14$. 
The NGB's ($\pi^a$) are contained in $\Sigma= e^{[i\pi^a X^a/f]} \langle \Sigma \rangle$, 
where $\langle \Sigma \rangle$ is the anti-symmetric condensate and $X^a$'s are the broken generators, 
as given in ~\cite{Low:2002ws}.
These Goldstone bosons obtain relatively small masses from the radiative corrections and 
therefore can be termed as pseudo-Nambu Goldstone bosons (pNGB). 
Eight of these pNGB's form two scalar doublets ($\phi_1$ and $\phi_2$), 
which, at the TeV scale, is similar to the scalar doublets of general 2HDM by construction. 
One major difference between a general 2HDM and the emergent 2HDM in the LSS model is the 
appearance of the vector like fermions and gauge bosons in the LSS model. 
These BSM particles along with the 2HDM structure in the LSS model can significantly 
alter the BSM scalar phenomenology, as shown in earlier works~\cite{Gopalakrishna:2015dkt,Gopalakrishna:2015wwa}.
In the following, we discuss the Yukawa couplings of the scalars and its effective 2HDM analysis. 
For a detailed description of the model we refer to~\cite{Low:2002ws,Gopalakrishna:2015dkt}. 

\subsection{Yukawa Sector: Diagonalization of Mass Matrices}
Diverse nature of the Yukawa sector in the LSS model requires special attention, as it has immense potential to
churn out intricacies of flavor and other EW physics. Here, construction of the
Yukawa sector involves an extension beyond the SM electroweak symmetry breaking.
In the LSS model, the $SU(6)$ symmetry is broken explicitly by the gauge and Yukawa couplings in multiple steps: termed as {\it collective symmetry breaking}.
In the fermion sector, this is ensured by a special structure of the Yukawa couplings~\cite{Low:2002ws}. 
In this model, we mainly focus on the couplings of the scalars with the third generation of fermions. 
Also, in this study we have adopted a different nomenclature for the fields  
compared to \cite{Gopalakrishna:2015dkt} and hence we show the following steps for clarity.
The Yukawa Lagrangian is, 
\bea
    {\cal L}_{Yuk} = y_1 f \bmat Q' & t^{\prime \prime} & (i\sigma^2 Q)^T & 0 \emat (\Sigma)^* \bmat 0 \\ t^c \emat
    + y_2 f \bmat 0 & 0 & Q^T & 0 \emat (\Sigma) \bmat i\sigma^2 Q^{\prime\, c} \\ t^{\prime \prime\, c} \\ 0 \\ b^{\prime \prime\, c} \emat \nonumber \\ -i y_{1b} f \bmat 0 & 0 & Q^T & 0 \emat (\Sigma) \bmat 0 \\ 0 \\ 0 \\ b^c \emat
    + i y_{2b} f \bmat 0 & 0 & (i \sigma_2 Q)^T & 0 \emat (\Sigma)^* \bmat 0 \\ b^c \\ 0 \\ 0 \emat 
    + \ . {\rm h.c.} \ .
\label{tYuk.EQ}
\eea
The new fermions are one vector-like quark doublet Weyl-fermion pair $Q'=(t',b')^T , {Q'}^c=(-{b'}^c,{t'}^c)^T $ with $Y=1/6$ and EM charge $2/3$,
one vector-like up-type quark singlet $t^{\prime \prime}, t^{\prime \prime\, c}$ with EM charge $\pm 2/3$, and one vector-like down-type quark singlet $b^{\prime \prime}, b^{\prime \prime\, c}$ with EM charge $\mp 1/3$. $Q=(t,b)^T$ is the SM SU(2) doublet. The $y_i$ and $y_{ib}$ (for $i=1,2$)
are the dimensionless couplings in the top and bottom sector respectively. 
Expanding the SU(2) structure of the Yukawa couplings and including the vector-like fermion masses we get,
\bea
    {\cal L}_{\rm mass + Yuk} \supset && - y_1  \left(f t^{\prime \prime} t^c - i {Q'}^T \phi_2^* t^c - i Q^T \cdot \phi_1 t^c \right)
    + y_2  \left(f Q^T \cdot {Q'}^c + i Q^T \phi_1^* b^{\prime \prime\, c} + i Q^T \phi_2^* t^{\prime \prime\, c} \right) \nonumber \\ 
    && + y_3 f ({Q'}^T \cdot {Q'}^c) + y_4 f  (t^{\prime \prime\, c} t^{\prime \prime}) + y_5 f (b^{\prime \prime\, c} b^{\prime \prime}) + y_{1b} (Q^T \phi_1^* b^c) - y_{2b} (Q^T.\phi_2 b^c) + h.c.. \ ,
    \label{Lferm.EQ}
\eea
Here $y_3$, $y_4$ and $y_5$ are also dimensionless constants.
From Eq.~(\ref{Lferm.EQ}), we can infer the fermion mass matrices after electroweak symmetry breaking (EWSB). The fermion mass matrices with EM charge $+2/3$ and $-1/3$ appears in the Lagrangian as,
\beq
    {\cal L} \supset \bmat t & t' &  t^{\prime \prime} \emat \bmat
    i y_1 \frac{v_1}{\sqrt{2}} & y_2 f &  iy_2 \frac{v_2}{\sqrt{2}} \\
     iy_1 \frac{v_2}{\sqrt{2}} & y_3 f & 0 \\
       -y_1 f & 0 &  y_4 f 
     \emat \bmat t^c \\ {t'}^c  \\  t^{\prime \prime\, c}\emat
     + \bmat b & b' & b^{\prime \prime} \emat
     \bmat y_{ib} \frac{v_i}{\sqrt{2}} & y_2 f & i y_2 \frac{v_1}{\sqrt{2}}  \\ 0 & y_3 f & 0 \\ 0 & 0 & y_5 f \emat
     \bmat b^c \\  {b'}^c \\  b^{\prime \prime\, c} \emat  + {\rm h.c.} \ ,
     \label{MudIntB.EQ}
\eeq
where $v_i = \{ v_1, v_2\}$, are $vev$'s of $\phi_1$ and $\phi_2$ respectively. 
After diagonalizing Eq.~(\ref{MudIntB.EQ}), we get couplings of the scalars to the fermions for the top sector, in their mass basis.
We implement a two-step diagonalization, where first the $f$-dependent terms are diagonalized analytically, and then the $v_{1,2}$ dependent terms (appear after EWSB) are diagonalized.
We define the rotations that diagonalize the $f$ dependent terms as,
\bea
\bmat t \\  t^\prime \\ t^{\prime \prime} \emat = \bmat c_{23} & -s_{23} &  0 \\ s_{23} & c_{23} & 0 \\ 0 & 0 & 1 \emat \bmat t_0 \\ t_1 \\ t_2 \emat ; \ 
\bmat t^c \\ {t^\prime}^c  \\ t^{\prime \prime\, c} \emat = \bmat c_{14} & 0 & -s_{14}  \\ 0 & 1 & 0   \\ s_{14} & 0 & c_{14} \emat \bmat t_0^c \\ t_1^c \\ t_2^c \emat ; \ 
\bmat b \\ b^\prime \\ b^{\prime \prime} \emat = \bmat c_{23} & -s_{23} &  0 \\ s_{23} & c_{23} & 0  \\ 0 & 0 & 1 \emat \bmat b_0 \\ b_1 \\  b_2 \emat, 
\label{fdiagRot.EQ}
\eea
with $s_{23} \equiv \sin\theta_{23} = y_2/(\sqrt{y_2^2 + y_3^2})$, $c_{23} \equiv \cos\theta_{23} = -y_3/(\sqrt{y_2^2 + y_3^2})$, and $s_{14} \equiv \sin\theta_{14} = y_1/(\sqrt{y_1^2 + y_4^2})$. 
After these rotations, some field redefinitions are performed to make the mass matrix entries real and positive.\footnote{ The field redefinitions are used, following the notation in Ref:~\cite{Gopalakrishna:2015dkt}:
$t_0^c = i \tilde t_0^c$, $t_1^c = -i \tilde t_1^c$, $t^{\prime\, c} = - \tilde t^{\prime\, c}$, $\psi_1 = i \tilde \psi_1$, and,
$b_1 = -\tilde b_1$, $\psi_2^c = -i\tilde \psi_2^c$, and $\psi_2 = i\tilde \psi_2$.}
For brevity of notation, tildes on the fields are dropped and we denote the fields $\tilde{\chi}_i$ simply as $\chi_i$ in the following. 
Hence the mass matrices in the Lagrangian take the form,
\beq
    {\cal L}^{\rm mass} \supset \bmat t_0 & t_1 & t_2 \emat
    \bmat {\cal M}^{t}_{11} & 0 & {\cal M}^{t}_{13} \\ {\cal M}^{t}_{21} & {\cal M}^{t}_{22} & {\cal M}^{t}_{23} \\ 0 &  0 & -{\cal M}^{t}_{33} \emat
    \bmat t_0^c \\ t_1^c \\ t_2^{c}   \emat 
    +  \bmat b_0 & b_1 & b_2 \emat
    \bmat {\cal M}^{b}_{11}  &  0 &  {\cal M}^{b}_{13} \\ {\cal M}^{b}_{21} & {\cal M}^{b}_{22} & {\cal M}^{b}_{23} \\  0  & 0 & {\cal M}^{b}_{33} \emat
    \bmat b_0^c \\ b_1^c \\ b_2^c   \emat
    + {\rm h.c.} \ .
    \label{Mfdiag.EQ}
\eeq
After the field redefinitions, the new mass matrix entries ${\cal M}^{t,b}_{ij}$ are,
\bea
{\cal M}^{t}_{11} &=& \frac{y_1 (y_3 y_4 v_1 + y_2 y_3 v_2 - y_2 y_4 v_2)}{\sqrt{y}_{14} \sqrt{y}_{23} \sqrt{2}} \ , \qquad
{\cal M}^{t}_{13} = \frac{(y_1^2 y_3 v_1 - y_2 y_3 y_4 v_2 - y_1^2 y_2 v_2)}{\sqrt{y}_{14} \sqrt{y}_{23} \sqrt{2}} \ , \nonumber \\
{\cal M}^{t}_{21} &=& \frac{y_1 (y_2 y_4 v_1 + y_2^2 v_2 + y_3 y_4 v_2)}{\sqrt{y}_{14} \sqrt{y}_{23} \sqrt{2}}  \ , \qquad \
{\cal M}^{t}_{23} = \frac{(y_1^2 y_2 v_1 - y_2^2 y_4 v_2 + y_1^2 y_3 v_2)}{\sqrt{y}_{14} \sqrt{y}_{23} \sqrt{2}} \ , \nonumber \\
{\cal M}^{t}_{33} &=& -f \sqrt{y}_{14} \ , \qquad   {\cal M}^{t}_{22} = f \sqrt{y}_{23} \ , \nonumber \\
{\cal M}^{b}_{11} &=& c_{23} y_{ib} \frac{v_i}{\sqrt{2}} \ , \qquad
{\cal M}^{b}_{13} = y_2 \frac{v_1}{\sqrt{2}} \ , \qquad
{\cal M}^{b}_{33} = y_5 f  \ , \nonumber \\
{\cal M}^{b}_{21} &=& y_{ib} \frac{v_i}{\sqrt{2}} s_{23} \ , \qquad 
{\cal M}^{b}_{23} = - y_2 \frac{v_2}{\sqrt{2}} s_{23},\
{\cal M}^{b}_{22} = \sqrt{y}_{23} f,
\label{Mvij.EQ}
\eea
where $\sqrt{y}_{14} \equiv \sqrt{y_1^2 + y_4^2}$, $\sqrt{y}_{23} \equiv \sqrt{y_2^2 + y_3^2}$.

In the next step, we diagonalize the $v_{1,2}$ dependent terms of the matrix ${\cal M}^t_{ij}$, where the 
entries of ${\cal M}^t_{ij}$ are functions of intrinsic parameters of the $SU(6)/Sp(6)$ LSS model. 
This is achieved through carrying out a bi-orthogonal transformation, constructing left and right orthogonal 
matrices $R_L$ and $R_R$ respectively, such that, 
\bea
R_L {\cal M}^t_{ij} R_R^T \equiv ({\cal M}^t_{ij})_D,
\eea
where $({\cal M}^t_{ij})_D$ is the diagonal matrix in the mass basis. 
Instead of showing explicit dependences of $R_L, R_R$, we perform the diagonalization along with the parameter space scan; where for each set of input parameters, the diagonalization process is being performed.
This process is iterated multiple times keeping the parameter dependence of these matrices.
The transformation of the fields are defined as,
\bea
(t_0\ t_1\ t_2)^T = R_L^T (T_2\ T_1\ t)^T \nonumber ,\\
(t_0^c\ t_1^c\ t_2^c)^T = R_R^T (T_2^c\ T_1^c\ t^c)^T,
\label{rot.EQ}
\eea
where $T_i, \ t$ and $T_i^c, \ t^c$ are the mass eigenstates in the top sector. 
We identify $(t, t^c)$ as the observed top-quark ($t$) mass eigenstates 
and the others correspond to the mass eigenstates of the vectorlike top partners ($T_1$, $T_2$).
In the bottom sector, the off-diagonal terms proportional to $v_i$'s are numerically insignificant 
due to the smallness of $y_{ib}$, and hence we do not perform the mass-diagonalization. 
We identify $(b_0, b_0^c)$ as the observed bottom quark ($b$) and the others correspond to vectorlike 
bottom partners ($B_1$,$B_2$).
We also identify the mass eigenvalues as $m_t, M_{T_1}, M_{T_2}$ in the top sector, 
and $m_b, M_{B_1}, M_{B_2}$ in the bottom sector. 
In order to identify the eigenvalues corresponding to $m_t, M_{T_1}, M_{T_2}$ and $m_b, M_{B_1}, M_{B_2}$, 
it is ensured that $m_t <M_{T_1}< M_{T_2}$ and $m_b<M_{B_1}<M_{B_2}$.
\subsection{Yukawa Couplings of the Scalars}
In this section, we are mainly interested in couplings of the SM like Higgs $h$ and the charged Higgs $H^{\pm}$ to the fermions because those couplings affect the flavor and EW observables the most as per the model parameter space of LSS model is concerned. We will discuss this further at the end of this section. In the following we present the structure of the relevant Yukawa couplings, i.e the couplings of $h$ and $H^{\pm}$ in detail. 

Before the mass diagonalization, the top-quark Yukawa coupling with Higgs-like scalar ($h$) in the flavor basis can be written as,

\beq
    {\cal L}_{htt} = \frac{i\, h}{\sqrt{2}} \bmat t & \psi_1 & t' \emat
    \bmat y_1 c_\alpha & -y_2 s_\alpha & 0 \\ 0 & 0 & 0 \\ -y_1 s_\alpha & 0 & 0 \emat
    \bmat t^c \\ \psi_1^c \\ t^{\prime\, c} \emat + {\rm h.c.} \ .
\eeq
Then we rewrite the Lagrangian in the basis where the $f$-terms are diagonal\footnote{In our convention,
we define the top-quark Yukawa coupling $y_{htt}$ as
$ {\cal L}_{htt} = (h/\sqrt{2}) y_{htt} \hat{t}_0 \hat{t}_0^c+ {\rm h.c.} $.
We define this with a positive sign here since our field redefinitions made the fermion mass terms positive.
}.
Also, we implement the field redefinitions as stated previously, to render the real fermion mass matrix.
Hence, the Yukawa coupling of $h$ in an intermediate basis now takes the form, 
\beq
{\cal L}^{\rm Yuk}_h \supset  \frac{h}{\sqrt{2}} \left[
  y_{00} t_0 t_0^c  + y_{01} t_0 t_1^c  + y_{10} t_1 t_0^c  + y_{11} t_1 t_1^c \right] + {\rm h.c.} \ ,
\label{hYuk.EQ}
\eeq
with 
$y_{00} \equiv (-y_1 c_\alpha c_{14} c_{23} + y_1 s_\alpha c_{14} s_{23} + y_2 s_\alpha s_{14} c_{23})$, 
$y_{01} \equiv (-y_1 c_\alpha s_{14} c_{23} + y_1 s_\alpha s_{14} s_{23} - y_2 s_\alpha c_{14} c_{23})$, 
$y_{10} \equiv (y_1 c_\alpha c_{14} s_{23} + y_1 s_\alpha c_{14} c_{23} - y_2 s_\alpha s_{14} s_{23})$,  
$y_{11} \equiv (y_1 c_\alpha s_{14} s_{23} + y_1 s_\alpha s_{14} c_{23} + y_2 s_\alpha c_{14} s_{23})$.
Yukawa couplings after the bi-orthogonal rotations to the mass-basis, as defined in Eq.(\ref{rot.EQ}) take the form,
\beq
 y_{htt} = \left[ y_{00} (R_L)_{31} (R_R)_{31} + y_{01} (R_L)_{31} (R_R)_{32} + y_{10} (R_L)_{32} (R_R)_{31} + y_{11} (R_L)_{32} (R_R)_{32} \right] \ ,
\eeq
where $(R_L)_{ij}$ and $(R_R)_{ij}$ with $i,j = \{1,2,3\}$, are the $(i,j)$-th entry of the rotation matrix $R_L$ and $R_R$ respectively.
We define the deviation of $htt$ from its SM value as,
\beq
      k_{tth} \equiv y_{htt}/y_{htt}^{SM}, 
\label{tth.EQ}
\eeq
where, $y_{htt}^{SM} = m_t/v$.
%
Similarly, Yukawa couplings of the CP even neutral scalar ($H$) can be obtained 
from Eq.(\ref{hYuk.EQ}) by making the change $c_\alpha \to s_\alpha$ 
and $s_\alpha \to -c_\alpha$. 
For the couplings of the pseudo scalar ($A$), we refer to~\cite{Gopalakrishna:2015dkt}.

The Yukawa coupling of the $b$-quark with $h$ and the mass of the $b$ quark can be derived from Eq.(\ref{Lferm.EQ}) as,
\beq
    {\cal L}_{hbb} \supset \frac{c_{23}}{\sqrt{2}} \left[v \left(y_{1b} s_\beta + y_{2b} c_\beta \right) + h \left(y_{1b} c_\alpha - y_{2b} s_\alpha \right) 
\right] b_0 b^c + {\rm h.c.} \ .
    \label{LhHAbb.EQ}
\eeq  
The coupling and the mass are thus identified to be;
\beq
m_b \equiv v (y_{1b} s_\beta + y_{2b} c_\beta )/\sqrt{2},~~~ y_{hbb}\sim m_b/v.
\eeq
Analogous mass and coupling expressions hold for the third generation lepton ($\tau$) with the replacement $y_b \to y_\tau$.

Similarly, couplings of the charged Higgs with the third generation quarks can be obtained as,
\beq
    {\cal L}^{Yuk}_{H^{\pm}} \supset H^+ \left( y^+_{00} b_0 t_0^c + y^+_{01} b_0 t_1^c + y^+_{10} b_1 t_0^c + 
    y^+_{11} b_1 t_1^c \right) +
      H^- \left(y^-_{00} t_0 b^c + y^-_{10} t_1 b^c + y^-_{02} t_0 \psi_2^c + y^-_{12} t_1 \psi_2^c  \right) + {\rm h.c.} \ ,
 \label{lHctb.EQ}
\eeq
where,
\bea
y^+_{00} &=& (y_1 s_\beta s_{23} c_{14} - y_1 c_\beta c_{23} c_{14} + y_2 s_\beta c_{23} s_{14}), \nonumber \\
y^+_{01} &=& (y_1 s_\beta s_{23} s_{14} - y_1 c_\beta c_{23} s_{14} - y_2 s_\beta c_{23} c_{14}),\nonumber  \\
y^+_{10} &=& (-y_1 s_\beta c_{23} c_{14} - y_1 c_\beta s_{23} c_{14} + y_2 s_\beta s_{23} s_{14}), \nonumber  \\
y^+_{11} &=& (-y_1 s_\beta c_{23} s_{14} - y_1 c_\beta s_{23} s_{14} - y_2 s_\beta s_{23} c_{14}), \nonumber  \\
y^-_{00} &=& [(-y_{1b} c_\beta + y_{2b} s_\beta) c_{23}], \hspace{0.2cm}
y^-_{10} = [(y_{1b} c_\beta - y_{2b} s_\beta) s_{23}], \nonumber \\
y^-_{02} &=& (-y_2 c_\beta c_{23}), \hspace{1.9cm}
y^-_{12} = (y_2 c_\beta s_{23}) \nonumber.
\label{YukCh.EQ}
\eea
After the numerical computation of the rotation matrices ($R_L, R_R$), that diagonalize the off-diagonal terms proportional to $v_{1,2}$, we obtain the Lagrangian involving the charged Higgs couplings with top and bottom quark as:
\beq
    {\cal L}_{H^{\pm}tb} \supset V_{tb} \frac{1}{v}  H^+ \bar{t} \left( y_{H^{\pm} t_L b_R} m_t P_R + y_{H^{\pm} t_R b_L} m_b P_L \right)  b  + {\rm h.c.} \ ,
 \label{lHctb.EQ}
\eeq
where the charged Higgs Yukawa couplings are given by,
\bea
 y_{H^{\pm} t_L b_R} = \left[ y^-_{00} (R_L)_{31}  + y^-_{10} (R_L)_{32} \right], \nonumber \\ 
 y_{H^{\pm} t_R b_L} = \left[ y^+_{00} (R_R)_{31}  + y^+_{01} (R_R)_{32} \right] .
 \label{YukCo.EQ}
\eea
The relevant Lagrangian involving the coupling of top and strange quarks with the charged Higgs can be obtained in a similar way, 
\beq
    {\cal L}_{H^{\pm}ts} \supset V_{ts} \frac{1}{v}  H^+ \bar{t} \left( y_{H^{\pm} t_L s_R} m_t P_R + y_{H^{\pm} t_R s_L} m_s P_L \right)  s  + {\rm h.c.} \ ,
 \label{lHctb1.EQ}
\eeq
where the $y_{H^{\pm} t_L s_R}, y_{H^{\pm} t_R s_L}$ can be derived from Eq.(\ref{YukCh.EQ}) with the 
replacement of $y_{1b} \to y_{1s}, y_{2b} \to y_{2s}$ and then use them in 
analogous to $y^-_{00}, y^-_{10}$ terms in Eq.(\ref{YukCo.EQ}). 
Note that, the first parts of both the couplings are similar and being proportional to $m_t$, they 
are the dominant terms. The second parts of the Lagrangian, being proportional to $m_b$ or $m_s$, 
are not significant due to smallness of the masses. Moreover, the couplings are of the same order, 
i.e, $y_{1b}\sim y_{1s}$, $y_{2b} \sim y_{2s}$. 
The coupling involving the down quark, $H^{\pm}td$, will also hold similar expression but the 
coupling is relatively small due to the smallness of $V_{td}$. Overall, the couplings 
$H^{\pm}td$ and $H^{\pm}ts$ will be smaller compared to $H^{\pm}tb$ 
\footnote {The couplings with lighter fermions that we neglect here, appears in squares in the flavor observables, 
and their effect is much smaller compared to the anomalies present in the flavor observables.}.  

Among the other couplings of the charged Higgs, the $H^\pm cs$ and $H^\pm \tau \nu$ couplings can be obtained as,
\beq
{\cal L}^{Yuk} \supset \left(y_{1c} c_\beta -  y_{2c} s_\beta \right) H^+ s c^c 
+ \left(- y_{1s} c_\beta +  y_{2s} s_\beta \right) H^- c s^c 
+ \left(- y_{1 \tau} c_\beta + y_{2 \tau} s_\beta \right) H^- \nu \tau^c + {\rm h.c.} \ .
\eeq
The charged Higgs decays dominantly through $H^{\pm} \to tb $ mode for most of the parameter space, 
unless there is an abrupt cancellation between different contributions of Eq.(\ref{lHctb.EQ})~\cite{Gopalakrishna:2015dkt}. 
Otherwise, the second most prominent decay mode $H^{\pm} \to cs $ takes over. 
The leptonic decay mode $H^{\pm} \to \tau^{\pm} \nu $ is not found to be very significant.
Hence the tree level charged current contributions from the diagrams containing $H^{\pm}$ is 
negligible in this model. 

The couplings of the neutral pseudoscalar 
($A$) and the heavy scalar ($H$) are studied in detail in Ref:~\cite{Gopalakrishna:2015dkt}, which shows that  
$H$ and $A$ couple to third generation quarks strongly. 
Their couplings to the leptons, as well as first and second generation of quarks are negligible. 
Hence, tree level FCNC processes in meson decays, neutral meson mixing and other lepton flavor violating decays are suppressed in this model. 
We discuss different structures of the Yukawa couplings in an effective 2HDM framework and their effects on FCNC's, in the next section.

\subsection{Effective 2HDM Framework of LSS model}
If the radiative corrections due to the presence of extra heavier gauge bosons, vectorlike quarks and the singlet scalars are included in the LSS model, an effective scalar potential at the TeV scale mimics that of the 2HDM, albeit with the absence of a number of scalar field combinations.
The scalar potential generated at one loop in the LSS model~\cite{Low:2002ws} is, 
\bea
{\cal V}_{LSS} = m_1^2 |\phi_1|^2 + m_2^2 |\phi_2|^2 + (b^2 \phi_1^T \cdot \phi_2 + {\rm h.c.}) + \lambda_5^\prime |\phi_1^T \cdot \phi_2|^2 \ , 
\label{2HDMeffL.EQ}
\eea
where $\phi_1$ and $\phi_2$ are the $SU(2)$ scalar doublets with hypercharge +1/2 and -1/2 respectively, and $\phi^T_1.\phi_2= \phi_1^T i\sigma^2\phi_2$ is the anti-symmetric product of the fields.
Similar to the 2HDM case, the ratio of the VEV's of the scalar doublets $\phi_{1,2}$ are presented in terms of
\beq
\tan\beta \equiv v_1/v_2 = \sqrt{m_2^2/m_1^2} \ ,
\label{tbeta.EQ}
\eeq
which is a function of the LSS scalar potential parameters $m_{i}^{2}$. 
The physical masses of the scalars that constitute the 2HDM structure can be expressed as, 
\bea
\label{mass-terms}
m_A^2 = 2 b^2/\sin{(2\beta)} \ ; \qquad m_{H^\pm}^2 = m_A^2 - \lambda_5^\prime v^2/2 \ , \nonumber \\ 
m^2_{H,h} = \frac{1}{2} \left[ m_A^2 \pm \sqrt{m_A^4 - 4 (m_A^2 -m_{H^\pm}^2) m_{H^\pm}^2 \sin^2{(2\beta)} }  \right] \ .
\eea
The parameters $m_1^2$, $m_2^2$, $b^2$ and $\lambda_5^\prime$ are functions of the $SU(6)/Sp(6)$ model Lagrangian parameters and can be expressed as,
\bea
\lambda_5^\prime &=& \frac{c g_1^2 \left[g_2^2 + (c'/c) y_2^2 \right])}{g_1^2 + g_2^2 + (c'/c)y_2^2} \ , \quad
b^2 = \frac{3 f^2}{8 \pi^2} y_1^2 y_2 (y_3 - y_4) \log{\frac{\Lambda^2}{M_f^2}} \ , \nonumber \\
m^2_{1\, f} &=& \frac{3 f^2}{8 \pi^2} (y_1^2 - y_2^2) (y_3^2 - y_4^2) \log{\frac{\Lambda^2}{M_f^2}} \ , \quad
m^2_{2\, f} = \frac{3 f^2}{8 \pi^2} (y_1^2 y_2^2 + y_2^2 y_5^2 - y_2^2 y_3^2 - y_1^2 y_4^2) \log{\frac{\Lambda^2}{M_f^2}} \ , \nonumber \\
m^2_{1g} &=& m^2_{2g} = \frac{3}{64\pi^2} \left[ 3 g^2 M_g^2 \log{\frac{\Lambda^2}{M_g^2}} + g^{\prime\, 2} M_{g'}^2 \log{\frac{\Lambda^2}{M_{g'}^2}} \right] \ , \quad 
m^2_{1s} = m^2_{2s} = \frac{\lambda_5^\prime}{16 \pi^2} M_s^2 \log{\frac{\Lambda^2}{M_S^2}} \ . 
\label{2HDMfLSS.EQ}
\eea
$\lambda_5^\prime$ gets gauge (proportional to $g_1$ and $g_2$) as well as fermionic contribution (proportional to $y_2$), 
whereas $b^2$ receives only fermionic 
contribution (proportional to $y_i$). In Eq.~\ref{2HDMfLSS.EQ}, $m_1^2$ and $m_2^2$ get contributions from the scalars (s), gauge bosons (g) and fermions (f) in the loop. The degrees of constructive or destructive interference depend on the sign of the coefficients.
$\Lambda$ is the cut-off which is taken to be $4\pi f$, where $f$ is the energy scale associated with $SU(6)/Sp(6)$ breaking. $M_f$ is the heavy vector-like fermion mass-scale. 
The heavy gauge boson masses are $M_g = f\sqrt{(g^2_1 + g^2_2)/2}$ and $M_{g^\prime} = f\sqrt{(g^{\prime\, 2}_1 + g^{\prime\, 2}_2)/2}$.  
The singlet scalar mass is $M_s = f \sqrt{c(g_1^2 + g_2^2)+c' y_2^2}$,
where $c$ and $c'$ are $O(1)$ parameters that depend on the details of UV completion, as explained in Ref:\cite{Gopalakrishna:2015dkt}. 

How the two scalar doublets $\phi_1$ and $\phi_2$ couple to other lighter fermions (lighter than the top) in the 2HDM, determines presence or absence of FCNC's in the theory. 
As tree level FCNC's are tightly bound by experimental absence, they can place stringent constraints on the model. 
The tree level FCNC's are absent in this emergent 2HDM from the LSS model, if the $Z_2$ symmetry is not broken. 
Symmetries in the LSS model compel top-quark to couple to both $\phi_1$ and $\phi_2$ (see Eq.(\ref{tYuk.EQ})), 
which breaks the $Z_2$ symmetry of the 2HDM in the top sector. 
This type of models with Type III 2HDM flavor structure results in non-trivial FCNC's in the third generation.
 
In the light fermionic Yukawa sector, $Z_2$ symmetry can be enforced, i.e. either $\phi_1$ or 
$\phi_2$ couples to the fermions. 
With the Yukawa interactions $y_{1(b,\tau,c)} \neq 0$ and $y_{2(b,\tau,c)}= 0$, i.e. light down type 
fermions being coupled only to $\phi_1$, stringent constraints from the LHC such 
as $h\to b\bar b,\tau \bar \tau$ measurements become important. 
The alternate possibility is $y_{1(b,\tau,c)} = 0$ and $y_{2(b,\tau,c)} \neq 0$, which relaxes the earlier constraints. 
If this Yukawa structure is adopted for the up type light fermions as well, this will resemble a Type I 2HDM 
framework in the light fermion sector, while the top sector will break it.
The earlier framework $y_{1(b,\tau,c)}\neq 0$ and $y_{2(b,\tau,c)} = 0$, along with the up type coupling 
with $\phi_1$ is also a Type I framework for the light fermions, but that is not so tenable from the LHC constraints. 

On the other hand, an alternative Yukawa pattern can be explored where $y_1 \neq 0$ for the up-type light fermions, 
them being coupled only to $\phi_1$ and the down-type fermions only couple to $\phi_2$ i.e. $y_2\neq 0$ 
for the down type. The light fermion sector in this scenario will look like a Type II 2HDM set up. 
If the top couples to both $\phi_1$ and $\phi_2$, it breaks the Type II structure, as seen earlier. 
The constraints from the $h\to b\bar b,\tau\bar\tau$ at the LHC are relaxed in this scenario. 
Hence these type of peculiarity in the flavor structure are potentially ripe for non trivial 
implications in the flavor physics sector. 
A detailed analysis of the impact of few hitherto important flavor observables in the 2HDM sector is the 
core of this work.

\section{Constraints from Flavor and EW physics}
\label{constrains}
The LSS model can exhibit 
different Yukawa patterns, leading to different 2HDM-like scenarios. 
Depending on the flavor structure of the Yukawa couplings, flavor changing-neutral 
currents can place important constraints on the model. 
We study the FCNC's involving different quarks, among which the third generation ones are in particular non-trivial.
In this model, the top-quark couples to both $\phi_1$ and $\phi_2$.
The Yukawa coupling involving the bottom-quark depends on the model parameters $y_{1b}$ and $y_{2b}$, 
which play a major role in the calculation of the flavor observables.
The choice of $y_{1b} \neq 0$, $y_{2b} \neq 0$ reflects the 
Type-III 2HDM-like scenario. 
We have also considered alternative scenarios when $y_{1b}= 0$, 
$y_{2b} \neq 0$ or $y_{1b}= 0$, $y_{2b} \neq 0$, which can 
lead to different 2HDM scenarios, as 
mentioned in section 2.3
\footnote{For a detailed study of the flavor structure of general 2HDM models, one might look at 
Ref~\cite{Crivellin:2013wna}.}.
Hence, in the LSS model we study three cases:
\begin{center}
Case I:~~$y_{1b} \neq 0$, $y_{2b} \neq 0$,\\
Case II:~~$y_{1b} = 0$, $y_{2b} \neq 0$,\\
Case III:~~$y_{1b} \neq 0$, $y_{2b}= 0$.\\
\end{center}
\subsection{Flavor Observables and $Zb\bar{b}$}

As reflected in the previous section, the effect of the Yukawa couplings involving the third generation 
quarks have dominant contribution in flavor and EW observables. 
Also, as mentioned earlier, the contribution of the tree level FCNC can be 
avoided in a large model parameter space of the LSS model. 
Hence, while discussing the FCNC's, we will be 
focusing only on the loop level FCNC's in the following text.
Note that, the vectorlike fermions also have significant impact on the flavor 
and EW observables~\cite{Ishiwata:2015cga}, but, in this note we focus on the scalar 
sector of the effective 2HDM in the LSS model.
We also include constraints from $Zb\bar{b}$ measurement, while the S and T parameters are kept 
well within limit by keeping the 
new gauge degrees of freedom in the heavier side ~\cite{Gopalakrishna:2015dkt}. 
We discuss different flavor observables dividing them into few broad categories.

 \paragraph {\underline {Radiative B-meson decays:}}
In context of the 2HDM, the most stringent constraint comes from the radiative $B$-meson decays, 
$B\rightarrow X_s\gamma$ ($B\rightarrow X_d\gamma$). 
The latest experimental and the theoretical value of BR($B\rightarrow X_s\gamma$) show discrepancy as 
indicated in Table~\ref{tab:1}. 
In the SM, the quark-level transition is mediated by $W$ boson and $t$ quark exchange via 
electromagnetic penguin diagram. 
The matrix element for this process at the electroweak scale is governed by the dipole operator.
The effective Hamiltonian for this process is given by,
\beq
\mathcal{H}_{eff}=-\frac {4 G_F}{\sqrt{2}}V^{*}_{ts} V_{tb}\sum_{i=1}^{8}  C_i(\mu) O_i(\mu).
\eeq 

Here, $V_{ij}$ represents the relevant Cabibbo-Kobayashi-Maskawa (CKM) factors. 
The $O_i$ are a complete set of renormalized dimension six operators. 
They consist of six four-quark operators, $O_{1-6}$, the electromagnetic dipole operator, $O_7$, 
and the chromo-magnetic dipole operator, $O_8$. These operators are evolved from 
the electroweak scale down to bottom mass scale 
using Renormalization Group Equations (RGE). 
The partial decay width of the quark level transition is given by,
\beq
\Gamma(b\to s\gamma) =\frac{\alpha G_F^2 m_b^5}{128 \pi^4}|V^{*}_{ts} V_{tb} C_7(\mu_b )|^2.
\eeq
The deviation in the observable is proportional to $\delta C_7(\mu_b)$, which 
can be expressed in terms of the effective Wilson coefficients at the matching scale $(\mu_W)$, 
given as $\delta C_7(\mu_W)$ and $\delta C_8(\mu_W)$. The explicit expressions for $C_{7,8}$ can be found in the literature 
\cite{Buchalla:1995vs}. 
In 2HDM like scenarios, such as in the LSS model, additional contribution
comes from the charged Higgs coupling to the quarks, given by
\beq
\mathcal{L}=\frac{\sqrt{2}}{v} H^{+} \bar t[V_{ij} m_{ui} \lambda_u^i P_L d_j + V_{ij} m_{dj} \lambda_d^j P_R d_j]+h.c.\ .
\eeq
Now, the contribution of the charged Higgs in $\delta C_{7,8}$ 
can be expressed as $\delta C_7(x)$ and $\delta C_8(x)$, computed as ~\cite{Misiak:2015xwa}~\cite{Hu:2016gpe},
\bea
\delta C_{7,8}=\left[{\lambda_{t}^2\over 3}F^{(1)}_{7,8}(x)-\lambda_{t}\lambda_{b} F^{(2)}_{7,8}(x)\right],
\label{delta}
\eea
where, $x = \frac{\bar m_{t}^2 (\mu_W)}{m_{H^{\pm}}^2}$, $\lambda_{t} \equiv y_{H^{\pm}t_{L}b_{R}}$ and $\lambda_{b}\equiv y_{H^{\pm}t_{R}b_{L}}$, as given in Eq.(\ref{YukCo.EQ}). 
In the following text, we choose a simpler notation and denote these effective Yukawa couplings as $y_{Ht_{L}b_{R}}$ and 
$y_{Ht_{R}b_{L}}$. 
The detailed expression of the function ``$F$'' is given in Ref~\cite{Ciuchini:1997xe}\cite{Mahmoudi:2009zx}.

The $B\rightarrow X_s\gamma$ branching ratio receives large contributions from 
the charged Higgs couplings with the top and bottom quark.
We impose the limits on the Wilson coefficient to be 
in the range $-0.063\le\delta C_7+0.242~\delta C_8\le 0.073~$.
The theoretical and experimental uncertainties are combined in a quadrature while deriving the limits.
It is interesting to note that the relative minus sign between the 
two contributions (from $W$ and $H^{\pm}$ in the loop) in general 2HDM model gives a destructive interference 
for some values of the model parameters~\cite{Mahmoudi:2009zx}. 
Whereas, in the LSS model, the coefficients ($\lambda_{t,b}$)
depend on the parameters of the strong sector and both destructive and constructive 
interference are possible depending on the values of the LSS model parameters. 
This is also true for the other observables as well.
Charged Higgs also contributes to $b\to d \gamma$ decay in the same manner. 
The coupling of the charged Higgs to the top and down quark is also similar to Eq.(\ref{lHctb.EQ}), 
but this coupling is suppressed in this model as $V_{ts} \gg V_{td}$.
 \paragraph {\underline {Neutral meson mixing:}}
In the SM, neutral meson mixing occurs due to the box diagrams with two $W$ exchanges in the loop level. 
In the case of $B_d$ and $B_s$ mesons, the hierarchical structure of the CKM matrix and the large mass of the 
top quark imply that the mixing is dominated by the diagrams involving the top quarks. 
In 2HDM like scenarios, the observables related to the neutral-meson mixing receive charged Higgs 
contributions~\cite{WahabElKaffas:2007xd}. 
Additional diagrams are included by replacing the $W$ lines with charged Higgs ones, 
yielding the contribution to the mixing as,
\beq
\Delta m_q = \frac{G_F^2}{24 \pi^2} (V_{tq}V^{*}_{tb})^2~\eta_B~m_B~m_t~f^2_{Bq}~I_{tot}(x_W,x_H,x_H),
\eeq
\beq
I_{\textrm{tot}}=I_{WW}(x_W)+ A_t^4 I_{HH}(x_H,x_H)
+2 A_t^2 I_{WH}(x_W,x_H),
\eeq

where, $x_W=\bar m_t^2/m_W^2$, $x_H=\bar m_t^2/m_H^2$. $I_{WW}$, 
$I_{HH}$ and $I_{WH}$ indicate respectively the internal bosonic 
lines of the corresponding diagrams with an external light quark 
$q=d,s$. For details of these Inami-Lim functions and other parameters we refer 
to Ref~\cite{Mahmoudi:2009zx,Abbott:1979dt}. We consider the charged 
Higgs contributions to the $B_s^0-\bar{B}_s^0$ mixing only, as the constraints are stronger 
than $B_d^0-\bar{B}_d^0$ mixing~\cite{Amhis:2016xyh}. 
Also, as mentioned earlier, the charged Higgs coupling to the top ($t$) 
and down ($d$) quark is suppressed in our model. Both experimental measurements 
and the SM predictions for $B_s^0-\bar{B}_s^0$ mixing are given in Table:\ref{tab:1}.
We obtain the total contributions to the mass splitting from the $W^\pm$ bosons, 
Goldstones and the charged Higgs boson ($H^\pm$) through the box graphs. 
Normalizing $\Delta m_{B_s}$ with respect to its SM prediction, 
we obtain the allowed range at $2\sigma$ to be $0.675 \le{\Delta m_{B_s}\over \Delta m^{\textrm{SM}}_{B_s}}\le 1.265~$.
\begin{table}[!h]
  \begin{center}
    \caption{\label{tab:1} {\it Standard model vs. experimental values of the flavor observables.}}
\begin{tabular}{|c|c|c|}
\hline
\hline
Observables & SM value & Experimental value\\
\hline
 $\textrm{Br}(B\rightarrow X_s\gamma)$ &$(3.36\pm 0.23)\times 10^{-4}$~\cite{Amhis:2016xyh}&$(3.32\pm 0.16)\times 10^{-4}$~\cite{Misiak:2015xwa}\\
$\Delta m_{B_s}$& $(17.757\pm 0.021)~\textrm{ps}^{-1}$~\cite{Amhis:2016xyh}&$(18.3\pm 2.7)~\textrm{ps}^{-1}$~\cite{Amhis:2016xyh}\\
$R_b$&$0.21581\pm0.00011$\cite{Haller:2018nnx}&$0.21629\pm0.00066$~\cite{ALEPH:2005ab}\\
$B_s \to \mu\mu$ & $(3.65 \pm 0.23)\times 10^{-9}$~\cite{Bobeth:2013uxa} & $(3.0\pm 
0.6^{+0.3}_{-0.2})\times 10^{-9}$~\cite{Aaij:2017vad}\\
\hline										
\end{tabular}
\end{center}
\end{table}
\paragraph {\underline {$\mathbf {Z\rightarrow b\bar{b}}$ vertex ($\mathbf {R_b}$):}}
The $Z\rightarrow b\bar{b}$ vertex has provided
opportunities to search for new physics contributions, due to the 
heavy masses involved in the loop. 
The radiative corrections 
at the vertex might imply charged Higgs exchanges in
addition to the standard $W$ boson coupling with top and bottom.
Precision measurement of electroweak precision observable $Z\rightarrow b\bar{b}$ 
branching ratio is measured as a ratio,
\begin{equation}
R_b={\Gamma(Z\rightarrow b\bar{b})\over \Gamma(Z\rightarrow \rm hadrons)}~.
\end{equation}
The modifications in $R_b$ due to the charged Higgs contributions 
at one loop is given by ~\cite{Haber:1999zh,Logan:1999if},
\begin{eqnarray}
\delta R_b\simeq-0.7785~\delta g^L_\textrm{new}~.
\end{eqnarray}
Here $\delta g^L_\textrm{new}$ is the modification in 
the $Zb_L\bar{b}_L$ coupling, calculated from a combination of 
triangle graphs where $H^\pm$ and the charged Goldstones float inside the 
loop. We neglect the modification in the $Zb_R\bar{b}_R$ coupling because 
it is proportional to $m_b^2$, compared to $Zb_L\bar{b}_L$ coupling which 
is proportional to $m_t^2$ ~\cite{Haber:1999zh}. 
We constrain $\delta R_b$ in the $2\sigma$ range, $-0.00086\le \delta R_b\le 0.00182~$.

\paragraph {\underline {Other flavor observables:}}
The flavor observables such as $B_s\to \mu\mu$ and $B_d\to \mu\mu$ can
also show sizable effect in generic 2HDM. 
But these observables will have 
low impact in the LSS model, as $H^{\pm}l\nu$ coupling is small, resulting in very tiny BR$(H^{\pm} \to l\nu)$ as  
discussed in the previous section. 
In general, standard 2HDM scenarios require very large values of the $H^+l\nu$ couplings 
with small charged Higgs mass~\cite{Crivellin:2013wna} in order to explain the current experimental 
values (Table\ref{tab:1}) of $B_{s,d}\to \mu\mu$. This is very unlikely to happen in the LSS model.
The other robust predictions from flavor observables such as $R_{D}$, $R_{D^*}$, $R_{K}$, $R_{K^*}$ and meson decays  
depend on BSM scalar couplings, $H^{\pm}l\nu$, 
$H^+H^-(Z/\gamma)$, $(H^0/h/A^0)l^+l^-$ ($l=\mu,\tau$), in both tree level and loop 
level contributions. 
These couplings, being small in the LSS scenario, give no effective contribution to the aforementioned  
flavor observables.
Hence, the constraining power of these observables are weaker, and we do not include them in our study.
\subsection{Constraints from flavor physics and $R_b$ {\bf (Scan A)}}
The flavor and electroweak observables depend on the following LSS model parameters,
\begin{center} 
$y_1$, $y_2$, $y_3$, $y_4$, $y_5$, $c$, $c^{\prime}$, $g_1$, $g_{1}'$, $y_{1b}$, $y_{2b}$, $f$ and $M$. 
\end{center}
The values of the top sector Yukawa couplings $y_i, i= 1,2..5$ are taken to be order of unity, which is a natural choice as 
they, as a combination, provide the top quark mass with VEV's $v_1, v_2 \sim 100~$GeV.
Three cases of the LSS model (case I, II and III) have different bottom Yukawa structure, depending on values  
Yukawa couplings $y_{1b}, y_{2b}$.  
The Yukawa couplings in the bottom sector, $y_{1b}$ and $y_{2b}$ are responsible for the bottom 
quark mass. 
To generate bottom mass, which is two orders of magnitude smaller than the top mass, $y_{1b}$ and $y_{2b}$
are varied in an optimum range, much smaller than the other Yukawa couplings. 
$y_{1b}$ and $y_{2b}$ are varied over a range $|y_{ib}| \le 0.10$. 
The intermediate Yukawa couplings $y_{00}^-, y_{10}^-$ have combinations of the form $y_{1b} \cos\beta - y_{2b} \sin \beta$, 
which chart out relative contributions of the $y_{1b}, y_{2b}$.
The rest of the parameters, $c$, $c_1$, $g_1$ and $g_{1}'$ 
are considered to be $\mathcal{O}$(1).

We have used the relation $M=1.5\times f$ where $f$ takes value upto 2 TeV 
amd we have fixed the VEV at 246 GeV. Hence, we perform a random multi-parameter scan in 12 parameter space. 
As noted in Ref:~\cite{Low:2002ws}, to prevent the VEV's running away to $\infty$, sufficient conditions  
are imposed, given as $m_{1,2}^2 > 0$, $(m_1^2 m_2^2 - b^4) < 0$ and $b^2$ is real. We have implemented these criteria 
in the multi-parameter scan as well.
As the flavor constraints are expected to be overwhelmed by the LHC constraints, 
we work in a more relaxed framework in terms of LHC parameters in order to  
emphasize the flavor intricacies of this type of set-up. 
The top and Higgs masses are fixed in the scan within a liberal window of 20~GeV and 10~GeV respectively
\footnote {Note that we impose a much stronger constraint on these parameters in the next scan, Scan {\bf B}.}.  
In order to see the effects emanating from the flavor sector, the flavor observables are kept within 
their current allowed limits as listed in Table.~\ref{tab:1}. Along with that, the bottom mass is also varied within 3-5 GeV.
We list the values of the parameters and constraints in the first scan, (Scan {\bf A}) in Table.~\ref{tab:2}.

\begin{table}[!h]
  \begin{center}
    \caption{\label{tab:2} {\it Values of the parameters and constraints from flavor observables used in the multiparameter Scan {\bf A}.}}
\begin{tabular}{|c|c|}
\hline
\hline
Parameter & Value\\
\hline
$y_i$(i=1....5), $c$, $c_1$, $g_1$, $g_{1}'$ & $\mathcal{O}$(1)\\
$y_{1b}$, $y_{2b}$& $\le 0.10$\\
$f$  & $\leq 2$ TeV\\ 
$M$  & $\leq 3$ TeV\\
VEV & 246 GeV\\
\hline
Quantity & Constraint\\
\hline
$\delta C_7+0.242~\delta C_8$ & ($-0.063,0.073$) \\
${\Delta m_{B_s}\over \Delta m^{\textrm{SM}}_{B_s}}$& ($0.675 , 1.265$)\\
$\delta R_b~$ & ($-0.00086, 0.00182$)\\
\hline										
\end{tabular}
\end{center}
\end{table}
\begin{table}[!h]
  \begin{center}
    \caption{\label{tab:3} {\it The constraints used in the multiparameter Scan {\bf B}.}}
\begin{tabular}{|c|c|}
\hline
\hline
Quantity & Constraint\\
\hline
$|\sin(\beta-\alpha)|$& $ \sim 1 $\\
$k_{tth}$ & $0.7-1.4$\\
$m_t$ & $156-170$ GeV\\
$m_h$ &  $123-127$ GeV \\
$m_b$ & $3-5$ GeV\\
$m_H^{\pm}$ & $> m_t$\\
$m_B$, $m_T$ & $> 1.4$ TeV \\
\hline										
\end{tabular}
\end{center}
\end{table}
\begin{figure}[tb]
\begin{center}
\includegraphics[width=6cm,height=6cm]{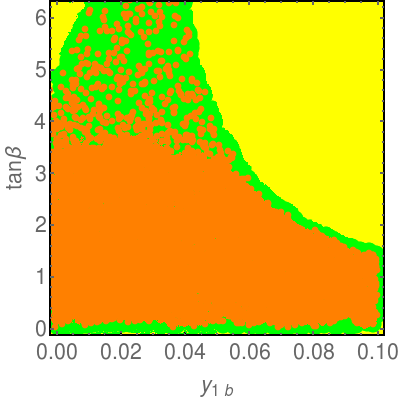}
\includegraphics[width=6cm,height=6cm]{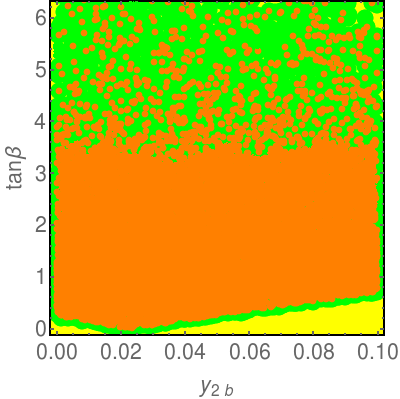}
\includegraphics[width=6cm,height=6cm]{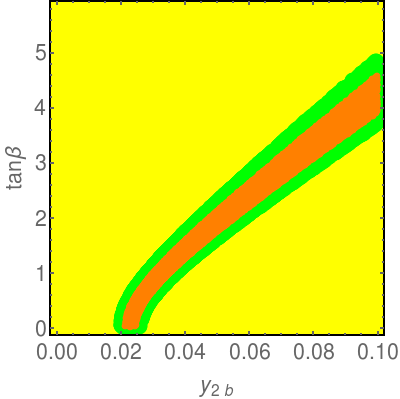}
\includegraphics[width=6cm,height=6cm]{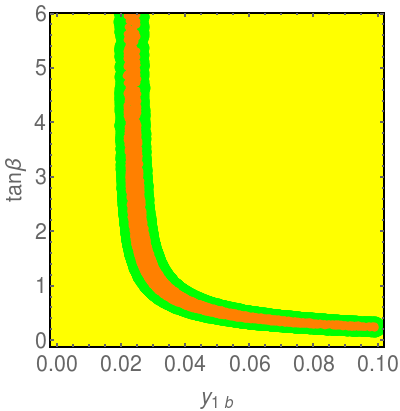}
\caption{\it Scan {\bf A}: Variation of $\tan\beta$ as a function of $y_{1b}$ and $y_{2b}$ in three cases, case-I (upper panel), case-II (lower panel left) and case-III (lower panel right), after imposing constraints from $R_b$ and $B_s-\bar{B_s}$ combined (green) and $B\rightarrow X_s\gamma$ (red).}
\label{fig:flavor1}
\end{center}
\end{figure}
\begin{figure}[tb]
\begin{center}
\includegraphics[width=4.9cm,height=4cm]{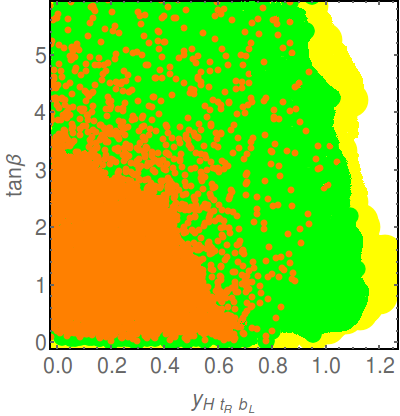}
\includegraphics[width=4.9cm,height=4cm]{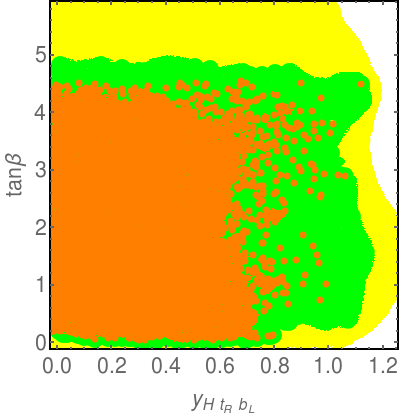}
\includegraphics[width=4.9cm,height=4cm]{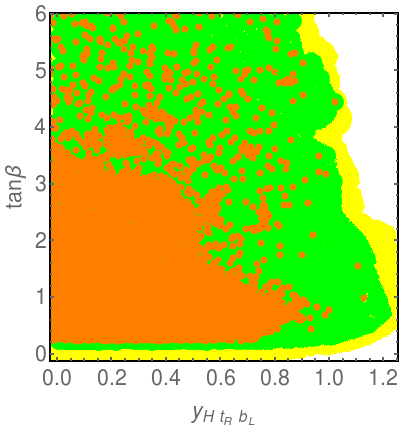}
\includegraphics[width=4.9cm,height=4cm]{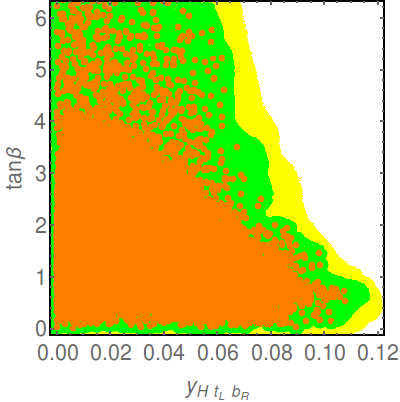}
\includegraphics[width=4.9cm,height=4cm]{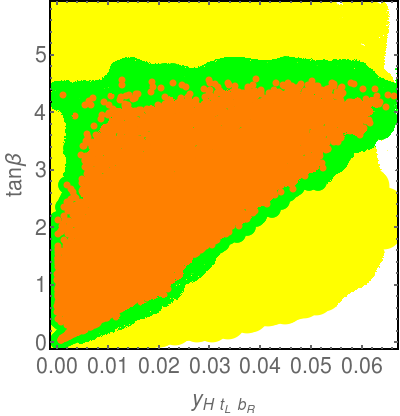}
\includegraphics[width=4.9cm,height=4cm]{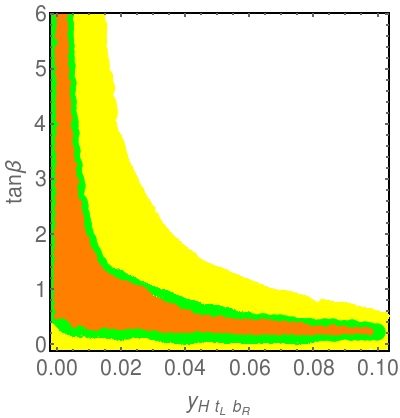}
\includegraphics[width=4.9cm,height=4cm]{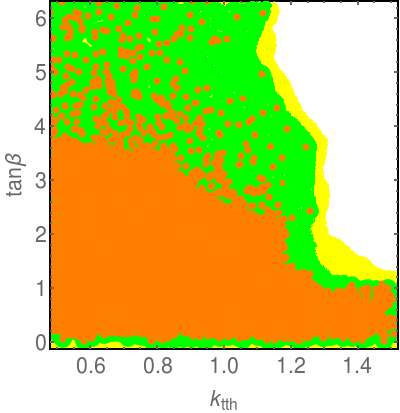}
\includegraphics[width=4.9cm,height=4cm]{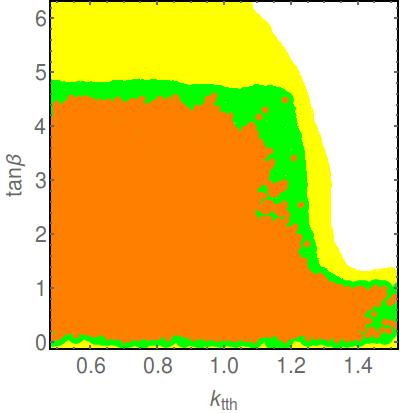}
\includegraphics[width=4.9cm,height=4cm]{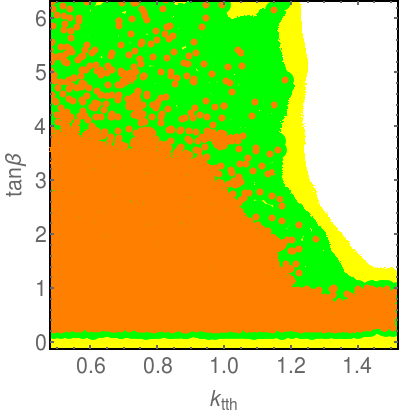}
\caption{\it Scan {\bf A}: Variation of $\tan\beta$ with the effective Yukawa couplings and $k_{tth}$ in three cases, case-I (left column), case-II (middle column) and case-III (right column), after imposing constraints from $R_b$ and $B_s-\bar{B_s}$ combined (green) and $B\rightarrow X_s\gamma$ (red).}
\label{fig:flavor2}
\end{center}
\end{figure}

In the following, we present Fig~\ref{fig:flavor1} and Fig~\ref{fig:flavor2} for three different cases, as specified earlier. In Fig~\ref{fig:flavor1}, we plot case I at the top panel and case II (left) and III (right) at the bottom panel. 
In Fig~\ref{fig:flavor2} we plot case I, II and III in the first, second and third column respectively. 
In these figures, the yellow points are generated by the model parameter space that is allowed after 
imposition of constraints on them, as given in the top part of Table.~\ref{tab:2}. 
With further imposition of constraints from the flavor and EW observables, green and red points 
depict the projection of allowed parameter region that satisfy the constraints (Table.~\ref{tab:2}) from $R_b$ and $B_s-\bar{B_s}$ mixing together and  
$B\rightarrow X_s\gamma$ respectively. 
It is found that $R_b$ and $B_s-\bar{B_s}$ satisfy almost the same parameter space. Hence we choose to impose 
combined constraints in the plots.
The flavor bound from $B\rightarrow X_s\gamma$ decay is found to be the most constraining for the LSS model. 
Hence, in the following discussion about the model parameter space constrained from flavor 
observables, we shall refer mainly to the strongest 
constraint that comes from $B\rightarrow X_s\gamma$.

We show the variation of the Yukawa parameters with $\tan\beta$ in Fig~\ref{fig:flavor1} and \ref{fig:flavor2}. 
In this model, both $m_1^2, m_2^2$ get contributions of same order of magnitude, through some combinations of the Yukawa couplings, $y_i, i =1,2...5$, which are of the same order. 
Hence, $\tan \beta$ in the LSS model, being written in terms of the ratio of $m_1^2, m_2^2$, is expected 
to lie in the vicinity of unity. It is reflected from the plots that overall small $\tan \beta$ values are preferred. 
As $\tan \beta$ values gradually increase from $<1$ to $>1$, upto $\tan \beta \sim 5$, 
relative dominance of the $\sin \beta$ and $\cos \beta$ dependent terms in Eq.(\ref{YukCh.EQ}) and (\ref{YukCo.EQ})
increase and decrease respectively. In Fig~\ref{fig:flavor1},
in the case-I scenario, where both $y_{ib} \neq 0$, flavor constraints seem to rule out the 
higher $\tan \beta$ region for higher $y_{1b}$ values along with 
lower $\tan \beta$ region for higher $y_{2b}$ values. 
This happens as the contributions to $y_{1b} \cos\beta - y_{2b} \sin \beta$ do not exactly cancel each other, 
thus paving way for increased contribution to the observables ($B\rightarrow X_s\gamma$ etc), 
constrained by experimental bounds. 

From Fig~\ref{fig:flavor1}, it is clear that the maximum value of $\tan\beta$ can be 5.0 for case-II, 
whereas for case-III, $\tan\beta$ upto 6 is mostly preferred by the flavor constraints \footnote {We do not show 
the region for $\tan\beta \geq 6$, as this region is less favored by the LSS model as well as the flavor 
observables, resulting in a very few points in the plot.} 
In case-II, for $y_{1b} =0$, the $\lambda_t$ values in Eq.(\ref{delta}) remain significant only for 
larger $y_{2b}$ and $\sin \beta$, which is related to $\tan \beta$. 
Only with large $y_{2b}$ and $\sin \beta$, $\lambda_t$ becomes larger which is paramount to fit the flavor data. 
Similarly, for case-III with $y_{2b}=0$, after a certain $y_{1b} (\ge 0.02)$, smaller values of $\tan \beta$
are preferred with increasing $y_{1b}$ as $\cos \beta$ becomes significant in that range and fails to provide 
the required values of $\lambda_t$ to satisfy the flavor constraints. Moreover, in case-III, $\tan \beta$ cannot 
obtain a very small value, $\tan\beta < 0.5$ region is disfavored from the flavor observables. 
Therefore, case-II with $y_{1b} = 0$ differs significantly from the case-I with all Yukawas non-zero, 
whereas case-III with $y_{2b} = 0$ remains relatively unaffected. 

The effective Yukawa pattern subsequently reveals the dominance of either $y_{1b}$ or $y_{2b}$ through their 
contributions in Eq.(\ref{YukCo.EQ}). 
From Fig:\ref{fig:flavor2}, we can see different ranges of $y_{Ht_Rb_L}$ and $y_{Ht_Lb_R}$. 
The allowed $y_{Ht_Rb_L}$ values are one order of magnitude higher than $y_{Ht_Lb_R}$ values, 
up to $\mathcal{O}$(1.0) as opposed to $\mathcal{O}$(10$^{-1}$). 
This is due to the fact that $y_{Ht_Rb_L}$ and 
$y_{Ht_Lb_R}$ are primarily dominated by the top sector and bottom 
sector Yukawa couplings ($y_i, i= 1,2$ vs. $y_{ib}, i=1,2$) respectively. 
They are also function of left and right top sector mixing elements in $R_L$ and $R_R$, which are of similar 
orders of magnitude, being function of $y_i$'s ($i= 1,2..5$). 
The effective Yukawa coupling $y_{Ht_Rb_L}$ is a combination of top sector Yukawas and there is no 
direct $y_{1b},y_{2b}$ contribution. Hence $y_{Ht_Rb_L}$ almost produces similar allowed regions 
after flavor constraints in 
three different cases. 
This is reflected in Fig.~\ref{fig:flavor2}, top row where we plot $y_{Ht_Rb_L}$ vs. $\tan\beta$. 
The flavor-allowed regions in case-III is somewhat similar to that of case-I, but case-II stands apart, 
where relatively larger values of $\tan \beta$ ($>4.5$) are not allowed.

The middle row of Fig.~\ref{fig:flavor2} shows the allowed regions of the effective Yukawa 
coupling $y_{Ht_Lb_R}$ with $\tan\beta$ for three cases. 
Large values of $y_{Ht_Lb_R}$ ($\sim 0.1$) are allowed for $\tan\beta< 1.5$ in 
case-I. 
Larger values of $\tan\beta$ ($\sim 4-5$) are favored in the region 
where $y_{Ht_Lb_R}$ is $>0.02$ in case-II whereas in case-III,  $\tan\beta\le 1$ is favored.
In case-III, smaller values of $y_{1b}$ ($<0.02$) prefers large $\tan\beta$. 
Another interesting feature of case-III is that a very small value of $\tan\beta$ ($<0.2$) is 
disfavored by $B\rightarrow X_s\gamma$, 
as shown in all the plots Fig.~\ref{fig:flavor2}(right column).

We also plot the $\tan\beta$ dependence of $k_{tth}$ in Fig:~\ref{fig:flavor2}. 
It is found that, in the experimentally allowed window of $0.7 < k_{tth} < 1.4$, strongly prefers $\tan\beta$ within 
the range $0.2-4.0$ in case-I and III, but values of $\tan\beta$ larger than 4 are also allowed by $B\rightarrow X_s\gamma$. 
In case-II, the upper limit on $\tan\beta$ is 4.0. 
The allowed values of the charged Higgs mass is almost same in these three cases. 
In the next section, we discuss the LHC constraints, starting from combined flavor constraints on the charged Higgs mass.
\subsection{Constraints from LHC {\bf (Scan B)}}
\begin{figure}[tb]
\begin{center}
\includegraphics[width=5.1cm,height=4cm]{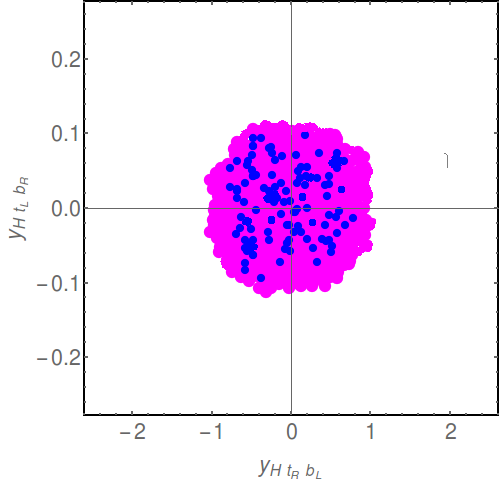}
\includegraphics[width=5.1cm,height=4cm]{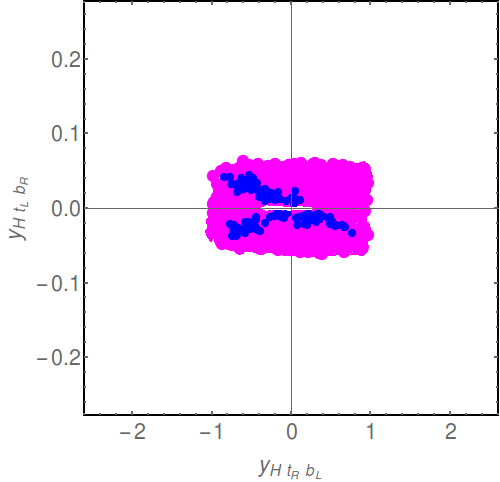}
\includegraphics[width=5.1cm,height=4cm]{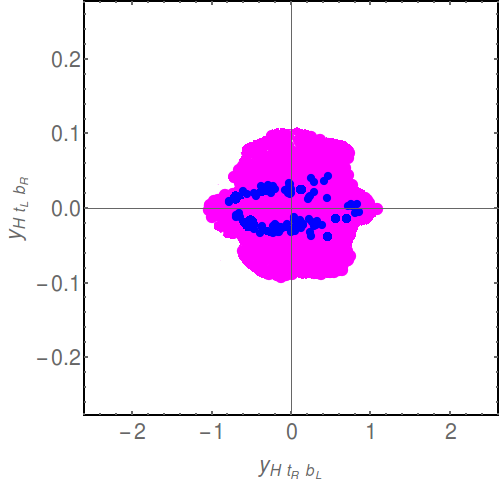}
\includegraphics[width=4.9cm,height=4cm]{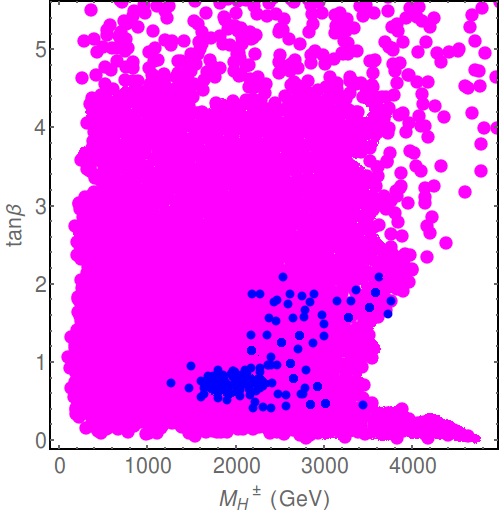}\hspace{0.2cm}
\includegraphics[width=4.9cm,height=4cm]{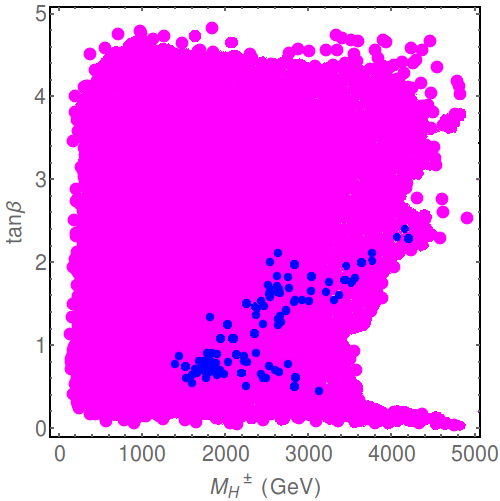}
\includegraphics[width=4.9cm,height=4cm]{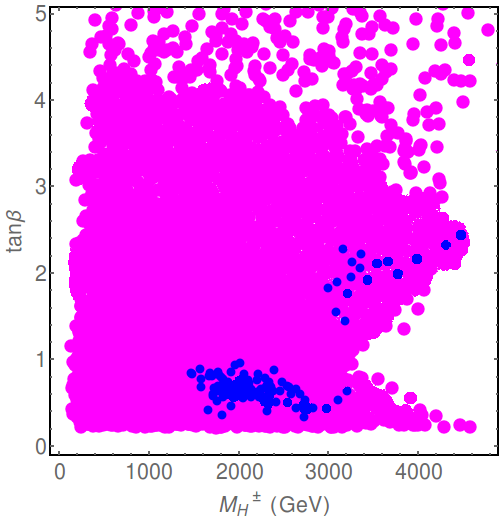}
\includegraphics[width=5.1cm,height=4cm]{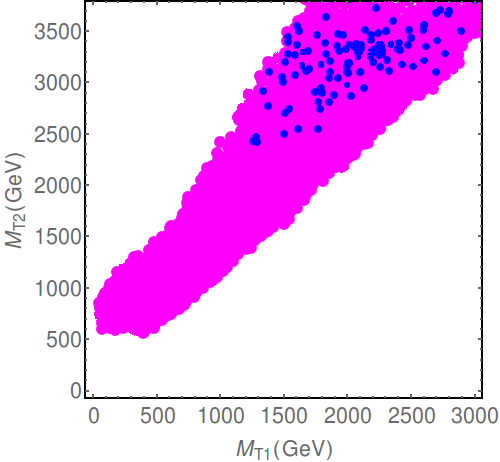}
\includegraphics[width=5.1cm,height=4cm]{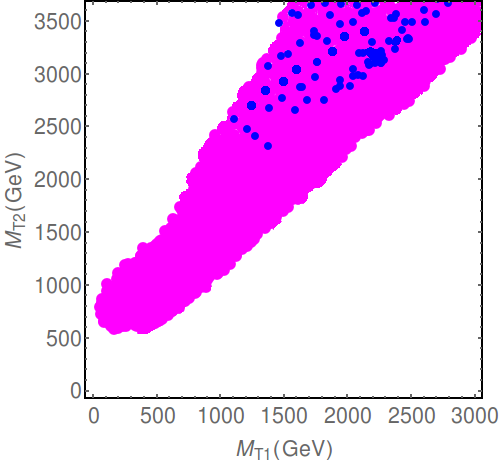}
\includegraphics[width=5.1cm,height=4cm]{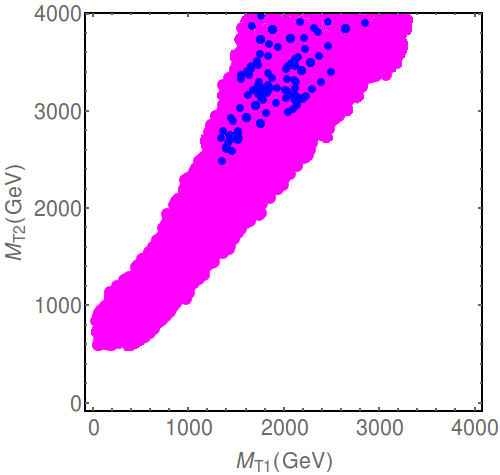}
\caption{{\it Scan {\bf B}:LSS model parameter space after imposing constraints from flavor observables and $R_b$ (Magenta) and further imposition of LHC observables (Blue) in three cases, case-I (left column), case-II (middle column) and  case-III (right column). 
}}
\label{fig:LHC0}
\end{center}
\end{figure}

At the LHC, properties of the Higgs boson are measured in several channels~\cite{Aad:2019mbh}, 
which indicates that the couplings of the 125 GeV Higgs to the gauge 
bosons ($hVV$) and to the fermions ($h f \bar f$) are consistent  
with the SM prediction ~\cite{Branco:2011iw,Chen:2018shg,Chen:2019pkq,Chen:2019rdk}, 
even though some mismatches are still there in the fermionic coupling measurements.
These measurements are found to be in favor of the alignment limit 
~\cite{Bhattacharyya:2015nca,Banerjee:2020tqc} in a 2HDM-like setup i.e. $|\sin(\beta-\alpha)| \sim 1 $.
The non observation of anomalous nature of the $hVV$ coupling strongly indicates that 
any BSM contribution is suppressed in the Higgs-gauge boson interaction, implying 
$|\cos(\beta-\alpha)|\sim 0$~\cite{Chen:2019bay,Su:2019ibd}. 
Hence, the $hVV$ coupling in the LSS model, which exhibits a 2HDM-like structure, are suppressed. 
Among the fermionic couplings, we only constrain the Higgs coupling to the third generation of quarks. 
We allow both positive and negative $hVV$ and $ht\bar t$ couplings 
leading to both constructive or destructive interference in the Higgs observables.
ATLAS and CMS experiments have imposed bounds on the couplings of the SM Higgs~\cite{Aad:2019mbh}, by looking 
at different production and decay modes of the Higgs boson. 
The current limit on $htt$ coupling is dependent on different assumptions held during the statistical fit, 
as shown in Ref:~\cite{Aad:2019mbh}. 
After inspecting the limits carefully, 
we perform our search in the window of $0.70 < k_{tth} < 1.4$, where $k_{tth}$ is defined as the 
ratio of the measured $htt$ coupling compared to its SM value (see Eq(\ref{tth.EQ})).
In the alignment limit,
the $hb\bar b$ coupling is $\sim m_b/v$, which is maintained throughout 
the multi-parameter scan.

We derive the masses of the SM and BSM particles in terms of the model parameters. 
We calculate the top mass in the $\bar{MS}$ scheme ~\cite{Alekhin:2012py,Tanabashi:2018oca} and constrain it in 
the range $156-170$ GeV around $m_t (\bar{MS}) = 163~$GeV. The Higgs mass 
~\cite{Aad:2012tfa,Chatrchyan:2012ufa} is considered in the range $123-127$ GeV.
LEP experiment excludes the charged Higgs masses below 80 GeV~\cite{Abbiendi:2013hk}. At the LHC, 
searches for the charged Higgs have been performed through various decay channels, 
$H^{\pm}\rightarrow cs$~\cite{Aad:2013hla,CMS-PAS-HIG-18-021}, $tb$~\cite{Aaboud:2018cwk,Sirunyan:2019arl} and 
$\nu \tau^{\pm}$~\cite{Sirunyan:2019hkq}, 
and most of these searches exclude $m_H^{\pm} < m_t$. However, the experimental 
searches~\cite{deFlorian:2016spz} in $H^{\pm}\rightarrow tb$ channel is 
favored at higher masses of $H^{\pm}$, as predicted in other 
BSM models, for example, in supersymmetric scenarios~\cite{Hicyilmaz:2017ntm}. 
Incidentally, also in the LSS model we get the $BR(H^{\pm}\rightarrow tb\sim 1)$ 
satisfying a large parameter space of the model.
Electroweak precision measurements~\cite{ALEPH:2005ab, Haller:2018nnx} 
require the charged Higgs mass to be close to the mass of one 
the neutral Higgses, which is also satisfied in our model (also 
shown in~\cite{Gopalakrishna:2015dkt}).
Different searches at the LHC have imposed limits on the masses of the vector 
like quarks as well, and typically they are in the range of 1 to 1.4 TeV ~\cite{Buckley:2020wzk}.
We list the values of all the constraining observables of the second Scan {\bf B} in Table.~\ref{tab:3}.
The model parameters follow the same values as given in Table.~\ref{tab:2}.  
\begin{table}[!h]
  \begin{center}
    \caption{\label{tab:3} {\it The experimental constraints, used in the multiparameter Scan \bf B.}}
\begin{tabular}{|c|c|}
\hline
\hline
Quantity & Constraints\\
\hline
$|\sin(\beta-\alpha)|$& $ \sim 1 $\\
$k_{tth}$ & $0.7-1.4$\\
$m_t$ & ($156,170$) GeV\\
$m_h$ &  ($123,127$) GeV \\
$m_b$ & ($3,5$) GeV\\
$m_H^{\pm}$ & $> m_t$\\
$m_B$, $m_T$ & $> 1.4$ TeV \\
\hline										
\end{tabular}
\end{center}
\end{table}

In Fig~\ref{fig:LHC0}, the magenta region shows the model parameter space allowed by combination 
of the flavor constraints and the $R_b$ (allowed points of Scan {\bf A}). 
We constrain the magenta region further with observables of the Scan {\bf B} as 
given in Table.~\ref{tab:3}. The result of Scan {\bf B} is shown by the blue points. 
The first row of Fig~\ref{fig:LHC0} shows the allowed parameter space of the Yukawa couplings, 
which contribute to the couplings of charged Higgs to the third generation quarks. 
Interestingly, after both the scans, we get very distinct patterns for the three cases.  
Parameter space that satisfies all the constraints (blue points) are uniformly distributed in case I. 
In case II, negative values of the $y_{Ht_Rb_L}$ are more preferred and
in case III, very small values of $y_{Ht_Lb_R}$ are not allowed after the imposition of LHC constraints.

In the second row of Fig~\ref{fig:LHC0} we show the plot of $\tan\beta$ vs. the mass of the 
charged Higgs boson. 
Mass of the charged Higgs 
is a function of, $\tan\beta$, $\lambda'_{5}$ and $b^2$, 
which 
depend intrinsically on other model parameters. 
The most important constraint 
form the LHC is on the $htt$ coupling, which again depends on the 
same set of model parameters.
The parameter space that survives after Scan {\bf B}, predicts 
such ranges of $m_H^{\pm}$ as shown by blue points in Fig~\ref{fig:LHC0}.
The preferred values of charged Higgs mass 
stay well above 1 TeV, the smallest allowed value 
being $\sim 1.3$ TeV, from Scan {\bf B}. 
Case-I shows more 
preference towards smaller values of $m_H^{\pm}$ compared to case-II.
Overall, the plots in the first and second row reflect 
mainly the correlation among the flavor physics observables, $R_b$ and Higgs Yukawa 
measurements at the LHC, in the allowed LSS model parameter space.

The bottom row confirms that the top-like VLQ partners are well above 1.3 TeV
after Scan {\bf B} and the allowed region is almost same in the three cases.
Our results also predict that the similar inference is true for 
the bottom-like VLQ's. 
In general, we find that the limits on VLQ masses and the charged Higgs mass 
are more stringent after the imposition of the LHC bounds, i.e, Scan {\bf B}, specifically due to the alignment condition.
Similar conclusion can be made for the extra neutral and charged gauge bosons, such as $B^{\prime}$ and $W^{\prime}$, where the 
allowed masses are pushed to even higher values, $\ge 2.2 ~$TeV. 
This also takes care of the constraints 
coming from electroweak precision tests (EWPT)~\cite{Gopalakrishna:2015dkt}.

\section{Conclusion}
\label{conclusion}
We re-introduce the LSS little Higgs model that acts as an UV complete model to produce an 
effective 2HDM at the EW scale, along with other BSM species, such as vector like fermions, 
extra gauge bosons and scalars. Compared to the generic 2HDM, emergent 2HDM from the LSS model will be predictive in nature, as the Yukawa structure here will be dictated by a bigger symmetry.
In this work, emergent 2HDM from the LSS model is discussed from the viewpoint of standard 2HDM to 
determine the crucial phenomenological differences. 
We bring out both flavor and LHC constraints together from the arsenal, ably aided by EW precision 
observable $R_{b}$ to pin down the allowed parameter space of the LSS model. 
In the context of an emergent 2HDM, qualitative features of an $SU(6)/Sp(6)$ composite Higgs model are expected to be similar to those for the LSS model, albeit with obvious difference in parametric dependences.

In the flavor sector, variants of models are constructed from the bottom Yukawa perspective, 
and we have found that the case with both $y_{ib} \neq 0$ is relatively less constrained compared 
to the cases with either of them put to zero. 
We deploy different flavor observables, potentially important to the charged Higgs sector. 
Among them the $B\rightarrow X_s\gamma$ branching ratio is found to be the most constraining, while $B_s-\bar{B}_s$ 
mixing and EWPT observable $R_b$ constraints are relatively liberate, though are not widely off. 
While the flavor constraints keep the charged Higgs fairly relaxed, the $\tan \beta$ gets more 
restricted ($\le 5$) compared to the usual 2HDM. 
The effective charged Higgs Yukawas, $y_{Ht_Lb_R} $ and $ y_{Ht_Rb_L}$ show very distinct 
$\tan \beta$ dependences in different cases (case-I,II,III), manifesting different flavor patterns.

With respect to the combined flavor and LHC constraints, charged Higgs mass ($m_H^{\pm}$) 
and $\tan \beta$ are bound tighter in the emergent 2HDM from the LSS model than in the usual 2HDM. 
Even when some of the constrained 2HDM scenarios rule out the charged Higgs only at sub-TeV region, here in the LSS model, 
combined flavor and LHC bounds push the charged Higgs mass lower bound to 1.3 TeV. 
Similarly, $\tan \beta$ is spread over a wider range in general 2HDM as opposed to a narrow 
range of (0.5-3) in the different LSS scenarios. 
This is a reflection of the predictive nature of the Yukawa 
sector in the LSS model, where the Higgs mass, top mass and top Yukawa couplings 
are fixed in terms of strong sector parameters of the LSS model. 
Whereas, these quantities were easily arranged in the construction of the general 2HDM 
through the enforcement of a $Z_2$ symmetry.

These type of LH models with a large number of parameters are often very fine tuned, as multiple parameters 
contribute to a single observable, such as the Higgs mass, which is very precisely known. 
It is worthwhile to know in future how the three different flavor scenarios, as discussed above, can have an impact 
on the charged Higgs phenomenology. 
In general, we find that the charged Higgs along with other neutral BSM particles are placed at masses well above one TeV. Hence, the decay products of them are expected to be highly energetic and that one can use to improvise different LHC search techniques. In one of the earlier LHC searches~\cite{Sirunyan:2020hwv}, heavy charged Higgs similar to this scenario which dominantly decays through $tb$ mode, was probed in all jet final states, using boosted properties of top and $W$. These kind of searches will have better prosepct at the advanced run of the LHC with enhanced luminosity and energy.


\section*{Acknowledgments} 
SS thanks UGC for the DS Kothari postdoctoral fellowship grant with award letter No.F.4-2/2006 (BSR)/PH/17-18/0126. N.K. acknowledges the support from the Dr.~D.~S.~Kothari Postdoctoral scheme (201819-PH/18-19/0013). 
SS thanks Dr. Shrihari Gopalakrishna for his guidance as PhD supervisor to study this model. 
SS also thanks Dr.~Shrihari Gopalakrishna and Dr.~Tuhinsubhra Mukherjee for their inputs on the LSS model construction and phenomenology during the course of an earlier work~\cite{Gopalakrishna:2015dkt}. We also sincerely thank the referee for a number of important, thought-provoking suggestions that have immensely helped us to improve our manuscript. 

\bibliographystyle{JHEPCust}
\bibliography{triplet_flavor.bib}
\end{document}